\documentclass{iopart}
\usepackage{iopams}
\usepackage{graphicx}
\usepackage{color}

\begin{document}
\title{Sensitivity Studies for Third-Generation Gravitational Wave Observatories}
\author{
S\,Hild$^{3}$,
M\,Abernathy$^{3}$,
F\,Acernese$^{4,5}$,
P\,Amaro-Seoane$^{33,46}$,
N\,Andersson$^{7}$,
K\,Arun$^{8}$,
F\,Barone$^{4,5}$,
B\,Barr$^{3}$,
M\,Barsuglia$^{9}$,
M\,Beker$^{45}$
% ...
}
\ead{stefan.hild@glasgow.ac.uk}
\vskip 1mm
% ...
\address{$^{3}$\,SUPA, School of Physics and Astronomy, The University of Glasgow, Glasgow, G12\,8QQ, UK}
\address{$^{4}$\,INFN, Sezione di Napoli, Italy}
\address{$^{5}$\,Universit\`{a} di Salerno, Fisciano, I-84084 Salerno, Italy}
% ...
\address{$^{7}$\,University of Southampton, Southampton SO17\,1BJ, UK}
\address{$^{8}$\,LAL, Universit\'{e} Paris-Sud, IN2P3/CNRS, F-91898 Orsay, France}
\address{$^{9}$\,AstroParticule et Cosmologie (APC), CNRS; Observatoire de Paris, Universit\'{e} Denis Diderot, Paris VII, France}
% ...
\address{$^{33}$\,Max Planck Institute for Gravitational Physics (Albert Einstein Institute) Am M\"{u}hlenberg 1, D-14476 Potsdam, Germany}
% ...
\address{$^{45}$ Nikhef, Science Park 105, 1098 XG Amsterdam, The Netherlands}
\address{$^{46}$ Institut de Ci{\`e}ncies de l'Espai (CSIC-IEEC), Campus UAB, Torre C-5, parells, $2^{\rm na}$ planta, ES-08193, Bellaterra,
Barcelona, Spain}

\begin{abstract}
Advanced gravitational wave detectors, currently under construction, are
expected to directly observe gravitational wave signals of astrophysical origin.
The Einstein Telescope, a third-generation gravitational wave detector, has been
proposed in order to fully open up the emerging field of gravitational wave
astronomy. In this article we describe sensitivity models for the Einstein
Telescope and investigate potential limits imposed by fundamental noise sources.
A special focus is set on evaluating the frequency band below 10\,Hz where
a complex mixture of seismic, gravity gradient, suspension thermal and radiation
pressure noise dominates. We develop the most accurate sensitivity model, referred to as ET-D,  for a
third-generation detector so far, including the most relevant fundamental noise 
contributions. 
\end{abstract}

\pacs{04.80.Nn, 95.75.Kk}

\section{Introduction}

The currently operating Gravitational Wave (GW) detectors LIGO \cite{ligo2}, Virgo \cite{virgo}, GEO\,600 \cite{geo} and
TAMA \cite{tama} are based on extremely sensitive Michelson interferometers. While the sensitivity achieved by these first 
generation detectors is mainly limited by
shot noise, mirror thermal noise and seismic noise, for the second generation
of instruments, such as Advanced LIGO \cite{aligo2}, Advanced Virgo \cite{adv}, GEO-HF \cite{Willke06} and LCGT \cite{lcgt06}, additional 
fundamental noise sources will start to play a role towards the low-frequency end of the
detection band: Thermal noise of the test mass suspension, photon radiation 
pressure noise and seismically driven gravity gradients acting on the test masses. 
These three sources of noise will become even more important for third-generation GW observatories such as the 
Einstein Telescope  (ET) \cite{et_punturo2010}, \cite{ET}, as these detectors aim to significantly increase the
detection band towards frequencies as low as a few Hz \cite{Hild08}, \cite{Hild10}.  Therefore, major parts
of the ET design are driven by exactly these noise sources. An overview of the
importance of the sub-10\,Hz band for astrophysical and cosmological analyses
can be found in \cite{et_punturo2010}. 

In this article we will give an overview of the currently ongoing ET design
activities with a special focus on the modelling of the achievable sensitivity
taking the most important fundamental noise sources into account. The first
sensitivity estimate for a third-generation interferometer was described in
\cite{Hild08}, \cite{FreiseGRG} and was based on a single interferometer covering the full frequency 
range from about 1\,Hz to 10\,kHz. In the following we will refer to this 
sensitivity curve as \emph{ET-B}. Subsequently we developed a more realistic
design, taking cross-compatibility aspects of the various involved technologies into account.
This led to the so-called xylophone-design, in which one GW detector is 
composed of two individual interferometers: A low-power, cryogenic low-frequency
interferometer and a high-power, room-temperature high-frequency interferometer. 
A detailed description of this xylophone detector sensitivity, in the following referred to as
\emph{ET-C}, can be found in \cite{Hild10}. The ET-C configuration will serve as
a starting point for the investigations described in this article. We improved
the sensitivity models for ET by including additional new noise sources as well
as by amending and updating noise contributions already previously included.
These improvements, which led to a new sensitivity estimate, referred to as
\emph{ET-D}, will be presented and discussed in this article.

In section \ref{sec:GGN} we discuss seismic and gravity gradient noise, followed by
the quantum noise contribution in section \ref{sec:QN}. Thermal noise of the
suspensions and test masses will be presented in section \ref{sec:TN}. An improved
noise budget for the Einstein Telescope is then given in section \ref{sec:ETD}. 
We conclude with a brief overview of the configuration of a full third-generation 
observatory, consisting of several GW detectors.  

\begin{figure}[Htb]
\centering
\includegraphics[width=1\textwidth]{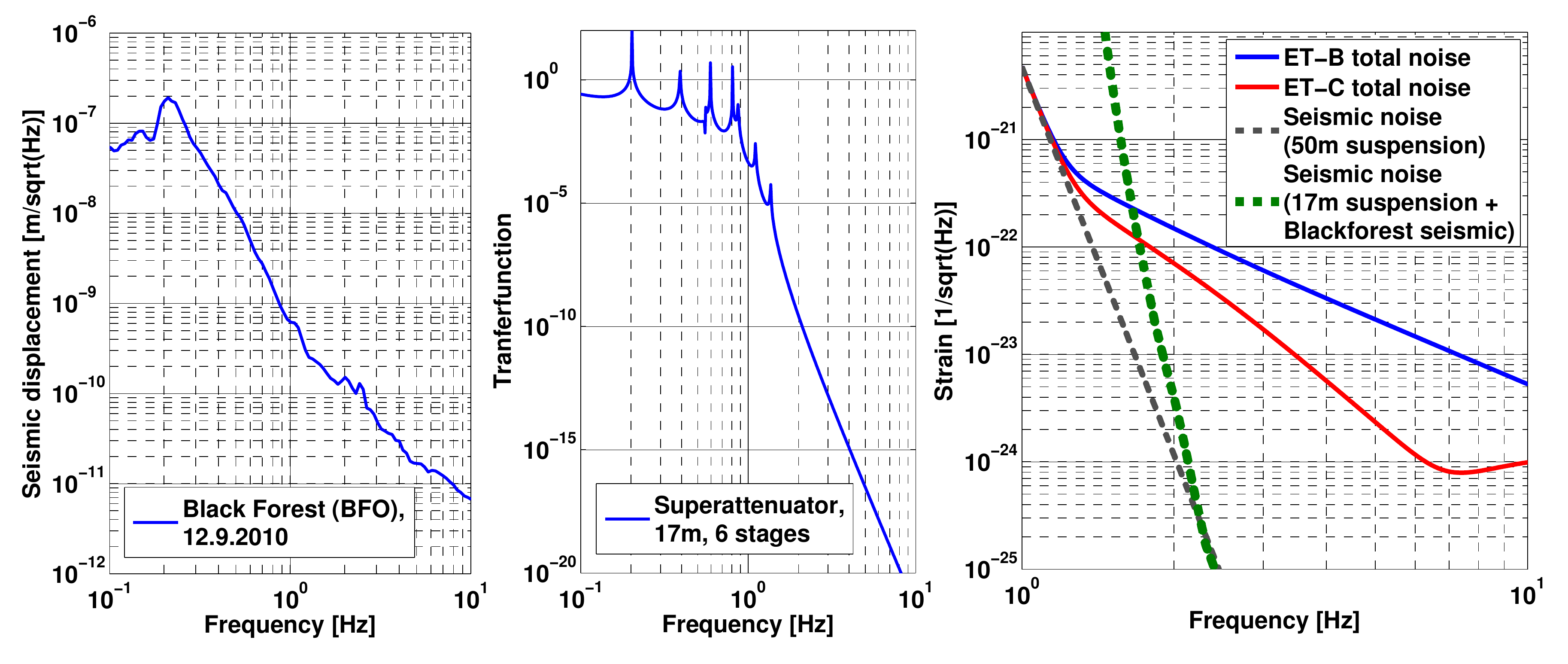}
\caption{Seismic noise spectrum from an underground location in the Black Forest, Germany 
(left hand panel). Transfer function of a superattenuator consisting of 6 stages with an
overall height of 17\,m (center panel). The right hand panel shows the resulting seismic
noise contribution for the 17\,m superattenuator for the seismic excitation at the 
Black Forest site (green dashed line). For comparison also ET-B and ET-C are plotted. Their seismic
noise contribution is based on the assumption of a generic 5-stage 50\,m suspension.} \label{fig:seismic}
\end{figure}

\section{Seismic Isolation and Gravity Gradient Noise}\label{sec:GGN}
Seismic noise couples into the differential arm length of a GW detector  
via two main paths. First of all, seismic excitation can mechanically couple through
the suspension and seismic isolation systems. Secondly, seismic noise excites density
fluctuations in the environment of the GW detector, which couple via
gravitational attraction to the test mass position. In the following we will refer to these two noise
sources as \emph{seismic noise} and \emph{gravity gradient noise}, respectively. 
The main difference between these two noise sources is that while seismic 
noise can be reduced by application of complex seismic isolation systems,
 the only guaranteed way to reduce the gravity gradient noise is to 
reduce the initial seismic excitation.\footnote{Many promising gravity gradient noise subtraction schemes have
been suggested in the literature \cite{BekerGRG}. However, as none of these schemes has been demonstrated 
so far, we do not consider them in this article.} Therefore, third-generation GW detectors are 
proposed to be built in quiet underground locations.

The seismic noise contribution of the low-frequency interferometer of ET-C was based on a seismic excitation of $5\cdot10^{-9}
{\rm m}/\sqrt{\rm Hz}/f^2$ (where $f$ is the frequency in Hz) and a generic 50\,m tall seismic isolation system consisting of 
5 passive pendulum
stages, each of 10\,m height. A more realistic seismic isolation design, based on the 
Virgo super attenuator concept \cite{Ballardin01}, \cite{Braccini05}, has been developed recently \cite{Braccini10}.
To achieve a lower cut-off frequency the height of the individual pendulum stages of the 
super attenuator will be extended to 2\,m per stage. The overall isolation of the 
proposed modified super attenuator, consisting of 6 pendulum stages (each stage providing 
horizontal as well as vertical isolation) and a total 
height of 17\,m, is shown in the center panel of Figure \ref{fig:seismic}. Using
the seismic excitation level, measured in an underground facility of the Black Forest
Observatory (BFO) \cite{BFO}, \cite{BFO_data}, shown in the left panel of Figure \ref{fig:seismic}, we
can derive the expected seismic noise contribution to the ET noise budget. The
result is shown in the right hand panel of Figure \ref{fig:seismic}. Reducing
the height of the seismic isolation system from 50 to 17\,m increases the 
 cut-off frequency only slightly from about 1.2 to 1.7\,Hz. 

Gravity gradient noise has been described in detail \cite{Saulson84, Beccaria98,
Hughes98}. In our simulations we estimate the power spectral density of the 
gravity gradient noise contribution as:
\begin{equation}
  N_{\rm GG}(f)^2 = \frac{4 \cdot \beta ^2 \cdot  G^2 \cdot \rho_r^2}{L^2 \cdot  f^4} \cdot  X_{\rm seis}^2
,
\end{equation}
where $G$ is the gravitational constant, $\rho_r$ the density of the rock around
the GW detector, $L$ the arm length of the interferometer, $f$ the frequency and
$X_{\rm seis}^2$ the power spectral density of the ground motion. $\beta$
accounts for the actual coupling transfer function from seismic excitation to
the differential arm length noise and depends for instance on the wave type of
the seismic excitation (e.g. ratio of P and S waves) and soil characteristics. 

\begin{figure}[Htb]
\centering
\includegraphics[width=0.47\textwidth]{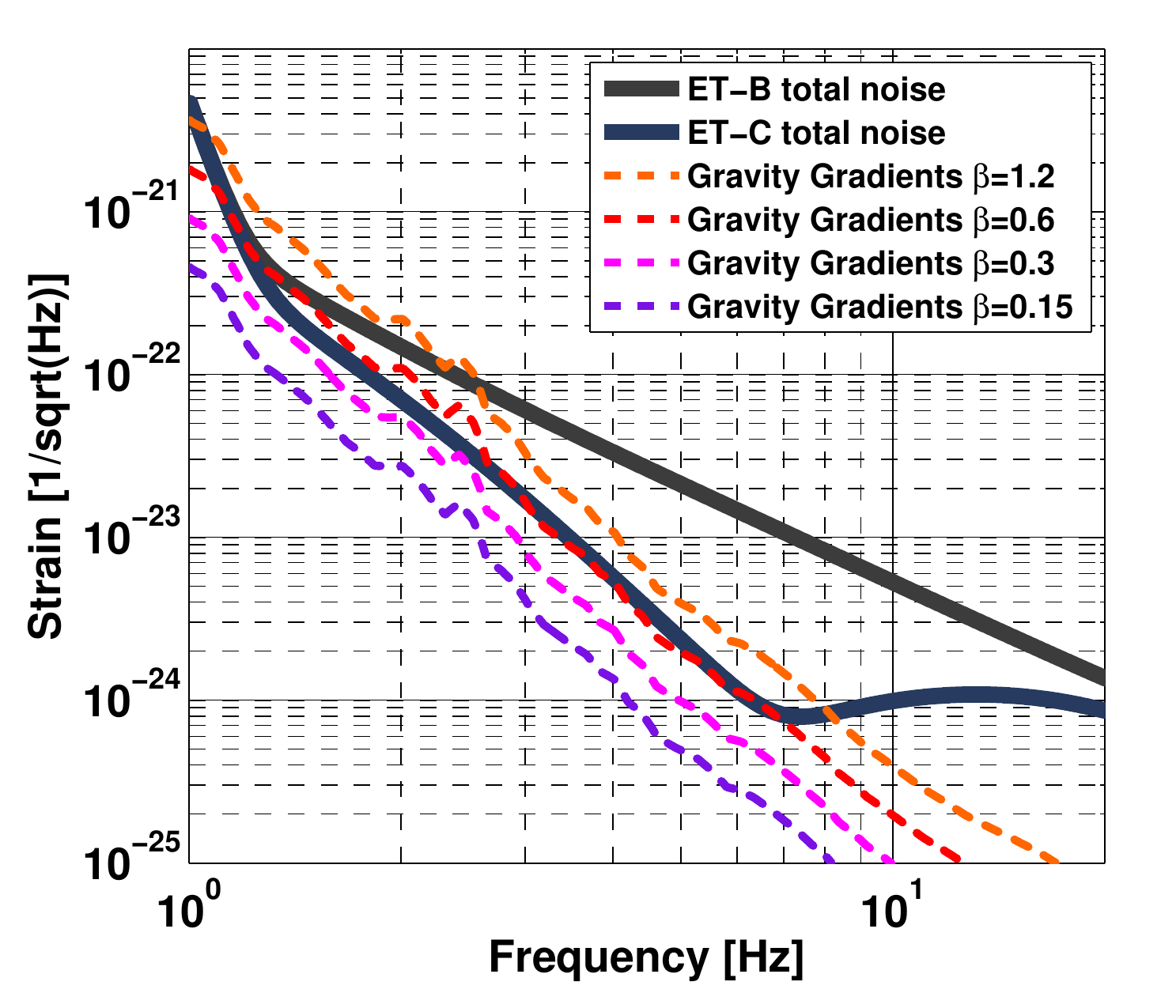}
\includegraphics[width=0.45\textwidth]{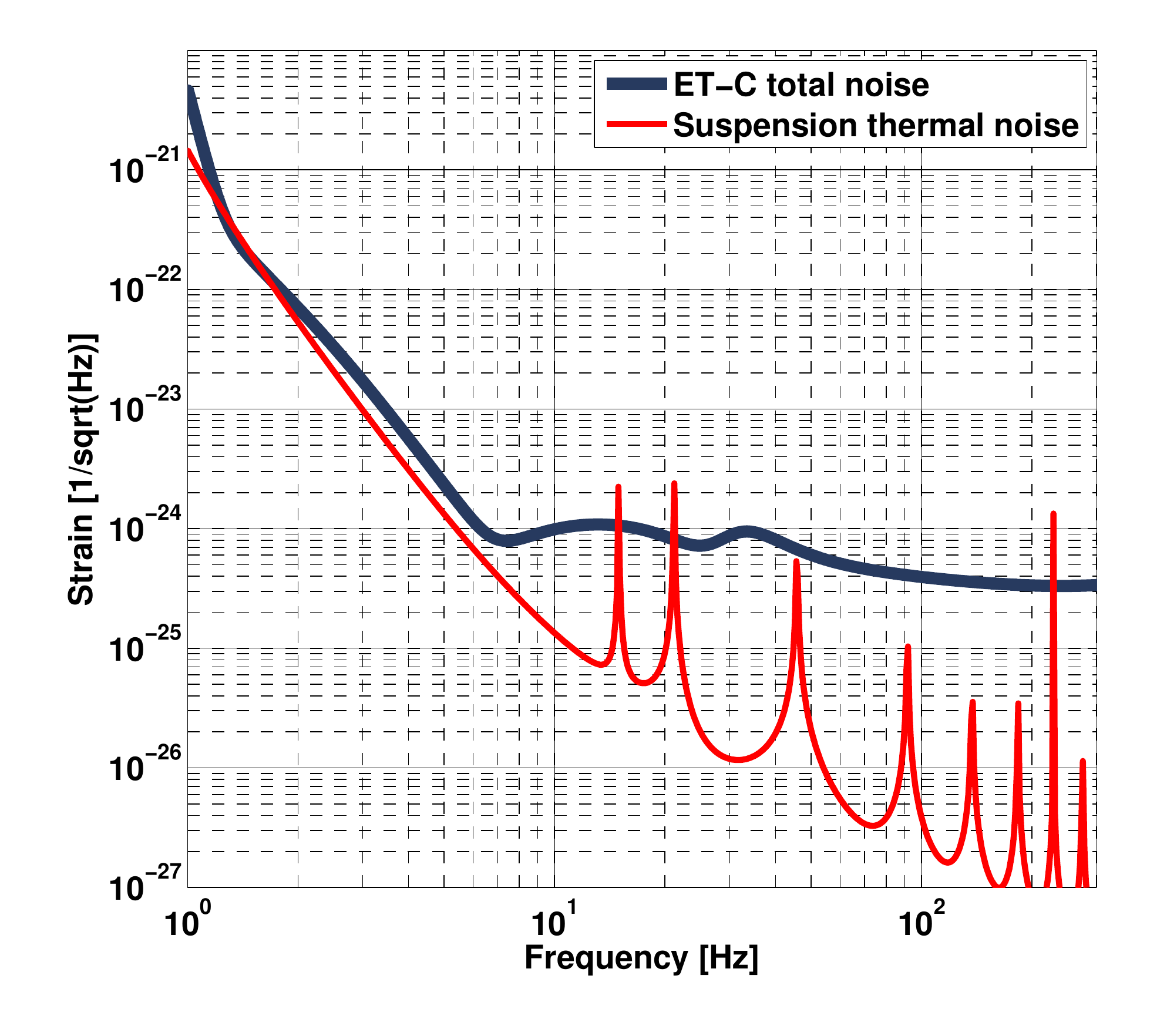}
\caption{Left panel: Gravity gradient noise contribution to ET, for various $\beta$ values, assuming the BFO spectrum
shown in Figure 1 as seismic excitation level. Right panel: Suspension thermal noise of the low frequency interferometer of ET as described in \cite{Ricci10}} 
\label{fig:GGN}
\end{figure}

Within the ET design study we carried out a campaign of measuring the seismic
noise in various underground locations across Europe. These measurements have
indicated that a couple of the test locations show a seismic excitation level
similar to or below the BFO measurement  \cite{Beker10}. Therefore, we assumed the BFO seismic 
excitation as a conservative estimate of a potential ET site.  The left panel of
Figure \ref{fig:GGN} shows the corresponding gravity gradient noise contribution
at the BFO site for different $\beta$. Since the detailed evaluation of a realistic $\beta$ for potential 
ET sites is an ongoing activity, we will use $\beta =0.58$, as given in the literature \cite{Saulson84, Beccaria98},
in the following for the ET-D sensitivity. Please note that our models do not
take atmospheric newtonian noise into account.

\section{Shaping of Quantum Noise}\label{sec:QN}

Quantum noise, composed of photon shot noise at high-frequencies and photon
radiation pressure noise at low frequencies, contributes significantly to the
overall sensitivity of ET's high frequency and low-frequency detectors. The 
high-frequency interferometers will feature a light power stored in the 
arm cavities of about 3\,MW to reduce shot noise, while the low-frequency
interferometers make use of only 18\,kW of light power in the arms, in order
to reduce the radiation pressure noise.
For ET-C
the quantum noise contribution was optimised by making use of detuned (low-frequency
interferometer) and tuned (high-frequency interferometer) signal recycling \cite{Meers88,
Hild07a}, together with an assumed generic quantum noise reduction of
10\,dB at all frequencies.

\begin{figure}[Htb]
\centering
\includegraphics[width=0.38\textwidth]{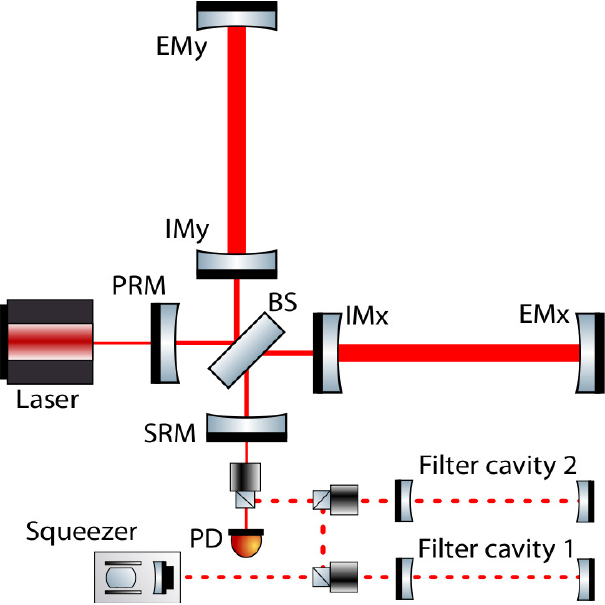}
\includegraphics[width=0.55\textwidth]{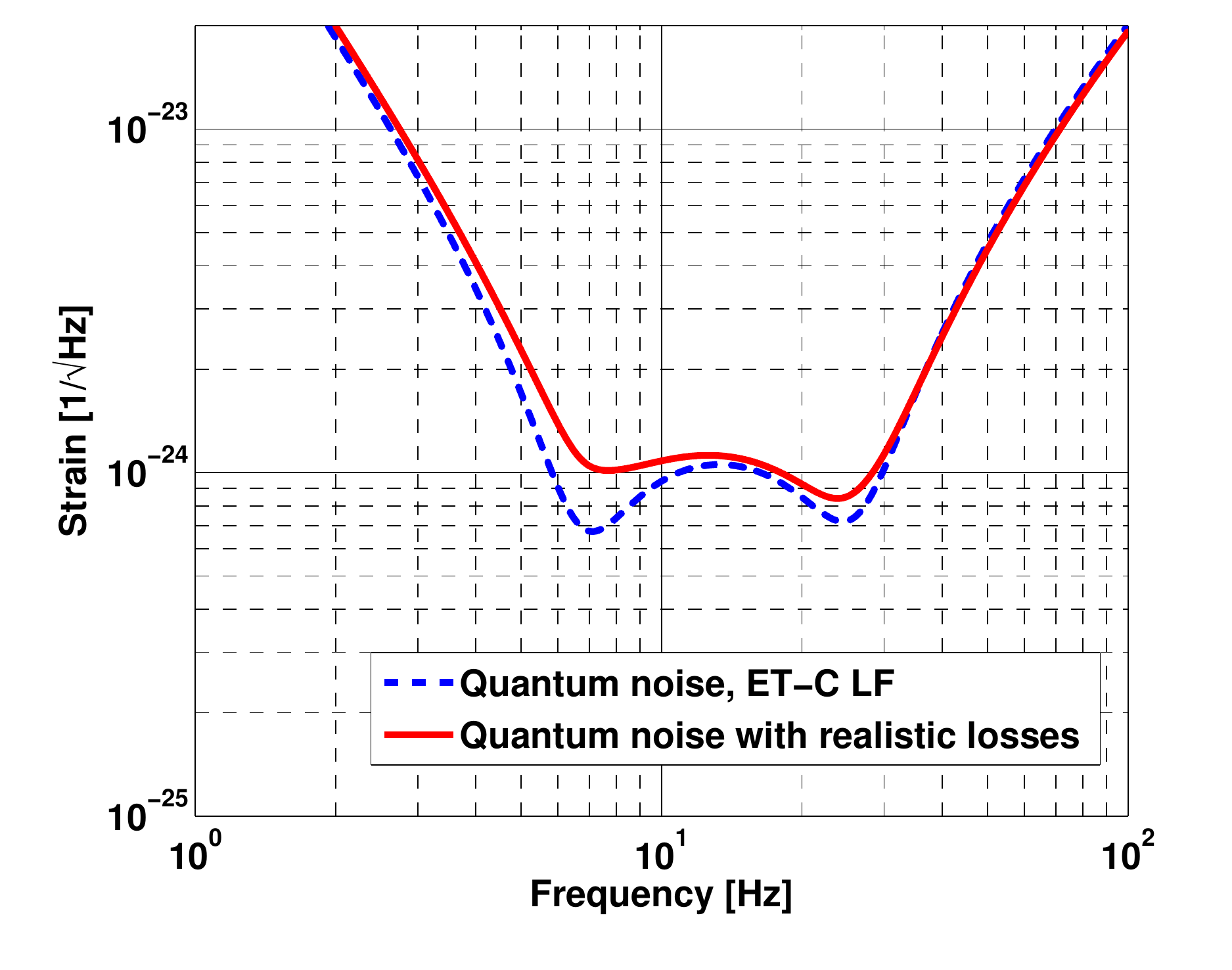}
\caption{Left panel: Simplified schematic of an ET interferometer. Quantum noise suppression is achieved
by the injection of squeezed light states with frequency dependent squeezing angle. The frequency dependent 
rotation of the squeezing angle can be realised by using the dispersion of filter cavities, on which the squeezed
light is reflected. Each ET low-frequency interferometer will require two filter cavities, while each high-frequency 
interferometer only requires a single filter cavity. Right hand panel: Quantum noise contribution
of the ET low-frequency interferometer, as described in \cite{Hild10} (dashed line) and with
squeezing losses from filter cavities taken into account (solid line) \cite{ET-0104A-10}.} \label{fig:QN_losses}
\end{figure}

Such a broadband quantum noise reduction can in principle be achieved by
injecting squeezed light states with a frequency dependent squeezing angle \cite{Kimble}
into the output port of the interferometer. Starting from a frequency independent 
squeezing angle, it is possible to use the dispersion occurring in reflection of a cavity, to create
squeezed light states with a frequency dependent squeezing angle. Figure \ref{fig:QN_losses}
shows a simplified schematic of a dual recycled interferometer with arm
cavities, consisting of a power recycling mirror (PRM), the beam splitter (BS),
the arm cavity mirrors (IM and EM) and the signal recycling mirror (SRM). In
addition the injection of the frequency dependent squeezed light states is also
shown: The squeezed light states leave the squeezing source (Squeezer) and are reflected at 
two filter cavities before they are injected via a Faraday rotator into the interferometer mode.
Finally, the interferometer output mode, consisting of the squeezed field and the signal field, 
is detected on the main photo diode (PD).   

In general, for an interferometer with signal recycling two filter cavities are 
necessary:  One for compensating the dispersion of the
 signal recycling resonance and one to reduce radiation pressure noise. The
bandwidth and detuning of the filter cavities depends on the actual optical
parameters of the main interferometers (e.g. arm length, SRM reflectivity, tuning 
of the signal recycling cavity) and need to be matched very accurately to
establish the full sensitivity improvement of the squeezed-light injection.
For the ET-D high-frequency interferometer we have the special case that 
only one filter cavity will be required, as it
employs tuned signal recycling with a bandwidth significantly larger than 
the cross-over frequency  of radiation pressure and shot noise.

The major loss mechanism for the squeezed light reflected off the filter cavities
originates from the fact that for frequencies close to the resonance of the
filter cavities, the squeezed states partly enter the filter cavity and
experience unavoidable roundtrip losses. We recently performed a detailed 
analysis of the requirements for the ET filter cavities as well as quantifying the 
squeezing losses inside the filter cavities \cite{ET-0104A-10}, \cite{Thuering10}.  
 We assumed a squeezing level of 10\,dB, an antisqueezing
level of 15\,dB and a loss of 75\,ppm per roundtrip inside a filter cavity. 
In order to reduce the influence of the intra-cavity losses, we chose a rather
long filter cavity length of 10\,km, which allows us to keep the filter cavity
finesse at a moderate level. The right hand
plot in Figure~\ref{fig:QN_losses} shows the corresponding quantum noise contributions of ET-D 
for the low frequency interferometer with filter cavity
losses included (solid lines).

\begin{figure}[Htb]
\centering
\includegraphics[width=0.51\textwidth]{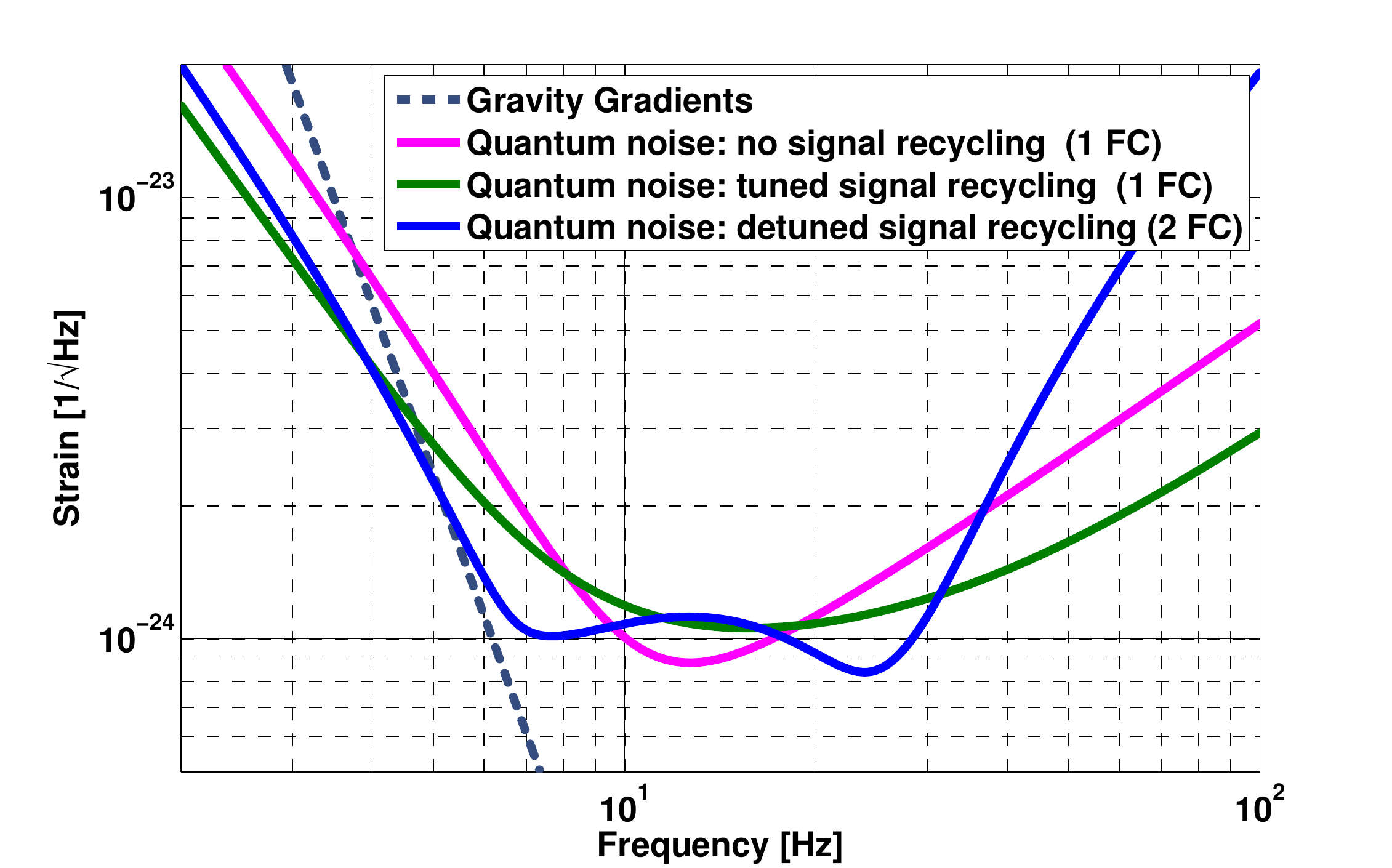}
\includegraphics[width=0.48\textwidth]{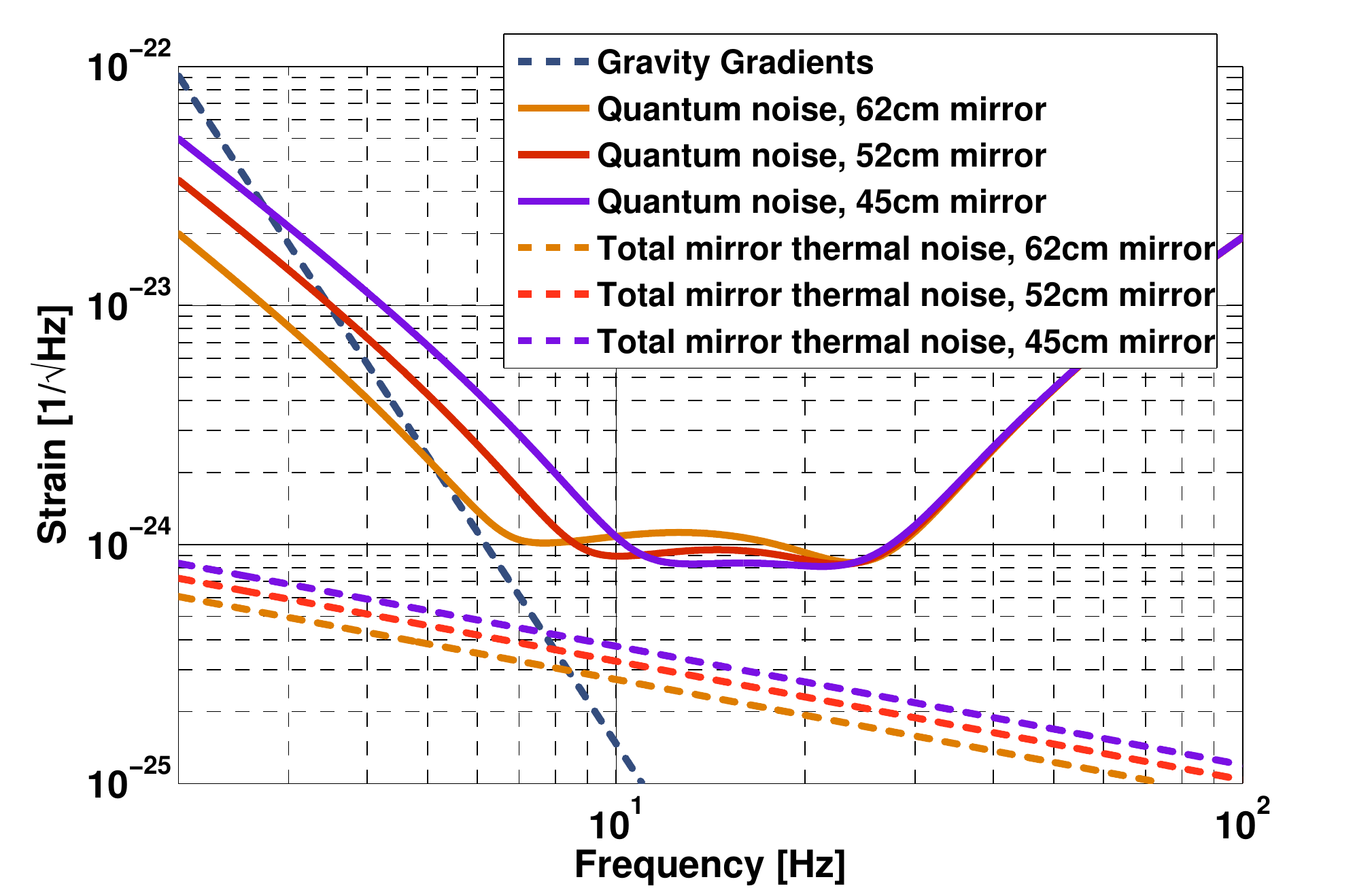}
\caption{Left: Quantum noise contribution for a low-frequency ET interferometer with different signal recycling options.
 For ET-D we assumed detuned signal recycling with SRM reflectivity of
80\,\%. Also plotted are a tuned signal recycling configuration using a 30\,\% reflectivity SRM and quantum noise without any signal recycling.
In brackets the number of required filter cavities is stated.  
 Right: Quantum noise and mirror thermal noise contributions for different mirror diameters. The aspect ratio is kept constant for all scenarios. Reducing
the mirror size (and thus their weight) only slightly increases the mirror thermal noise contributions, but significantly decreases the sensitivity at low 
frequencies, due to increased 
radiation pressure noise.} \label{fig:SR_options}
\end{figure}

For the low frequency interferometer it is also interesting to compare different
signal recycling options, properly accounting for the squeezing losses inside
the filter cavities. The left hand plot of Figure \ref{fig:SR_options} shows
the ET-D configuration, employing detuned signal recycling together
with two filter cavities, in comparison to tuned signal recycling as well as an
configuration without signal recycling. Both of these latter configurations
require only a single filter cavity, but this comes at the price of significantly lower 
sensitivity in the frequency band of interest (4--30\,Hz).  

\section{Thermal noise contributions}\label{sec:TN}

Brownian fluctuations couple into the differential arm length signal as
thermal noise of the test mass itself and of its suspension. Both of these
noise contributions can be significantly reduced by lowering the 
temperature of the test masses and the corresponding suspensions. 
The Japanese CLIO project \cite{CLIO04} has successfully demonstrated the operation 
of a laser interferometer at cryogenic temperatures. The recently funded 
LCGT detector \cite{lcgt06} is expected to transfer cryogenic technologies to 
a full scale second-generation GW detector. 

While for the ET high frequency interferometers even at room temperature the various thermal noise
contributions either do not play a significant role or can be sufficiently
reduced by increasing the beam size on the test masses and the use of 
higher-order Laguerre Gauss beam shapes \cite{Chelkowski09} or 
so-called `Mesa' beams \cite{Miller08}, the ET low-frequency interferometers
are expected to operate at cryogenic temperature.

 In the frequency band from 1
to 10\,Hz suspension thermal noise is the dominating thermal noise contribution.
When operating the low frequency interferometer at cryogenic temperatures the 
last stage suspension does not only need to be compliant with the thermal noise
requirements, but the actual design is also driven by the requirement to extract
any heat (deposited by the laser beams in the test masses) via the suspension. Our model \cite{Ricci10} assumes 
a mirror temperature of 10\,K, silicon fibres of 2\,m length and 3\,mm diameter as well as a temperature
of the penultimate mass of 2\,K.  
The righthand plot of Figure \ref{fig:GGN} shows the simulated suspension thermal noise
contribution for the ET low frequency detector \cite{Ricci10} using a branched
system of multiple oscillators consisting of the main mirror, the penultimate mass and the reaction 
mass \cite{Puppo10a}, \cite{Puppo10b}.

Fused silica, which is the material of choice for the test masses of all first generation 
GW detectors, has a high dissipation at low temperatures and therefore cannot be
used as substrate material for cryogenic test masses \cite{Andreson55}, \cite{Wider00}. Sapphire and silicon 
have been proposed as alternative materials \cite{lcgt06}, \cite{Rowan03} and there are strong R+D efforts to evaluate 
the optical and mechanical properties of these two candidate materials.  In the following
we assume silicon as test material for ET, but sapphire would yield similar results.  A detailed noise analysis 
of a cryogenic test mass for ET is given in \cite{ET-021-09}. The total 
thermal noise of the test masses is a combination of coating Brownian, substrate Brownian, substrate 
thermoeleastic and thermo-optic noise.  Of these four contributions coating Brownian noise is the most 
important one because in the frequency range of
interest it is a factor of about 5 larger than any of the other thermal noises.
The power spectral density of coating brownian noise can be described as
\begin{equation}\label{eqn:coat-noise}
S_x(f) = \frac{4 k_{\rm B} T}{\pi^2 f Y }\frac{d}{r^2_0}\left( \frac{Y'}{Y}\phi_{\parallel}+\frac{Y}{Y'}\phi_{\perp}\right)
\end{equation} 
where $f$ is the frequency, $d$ the total thickness of the coating, $r_0$ describes
 the beam radius, $Y$ and $Y'$ are the Young's modulus values for the substrate 
and coating respectively. $\phi_{\parallel}$ and $\phi_{\perp}$ are the mechanical
loss values for the coating for strains parallel and perpendicular to the coating
surface \cite{Harry02}.

\begin{figure}[Htb]
\centering
\includegraphics[width=0.48\textwidth]{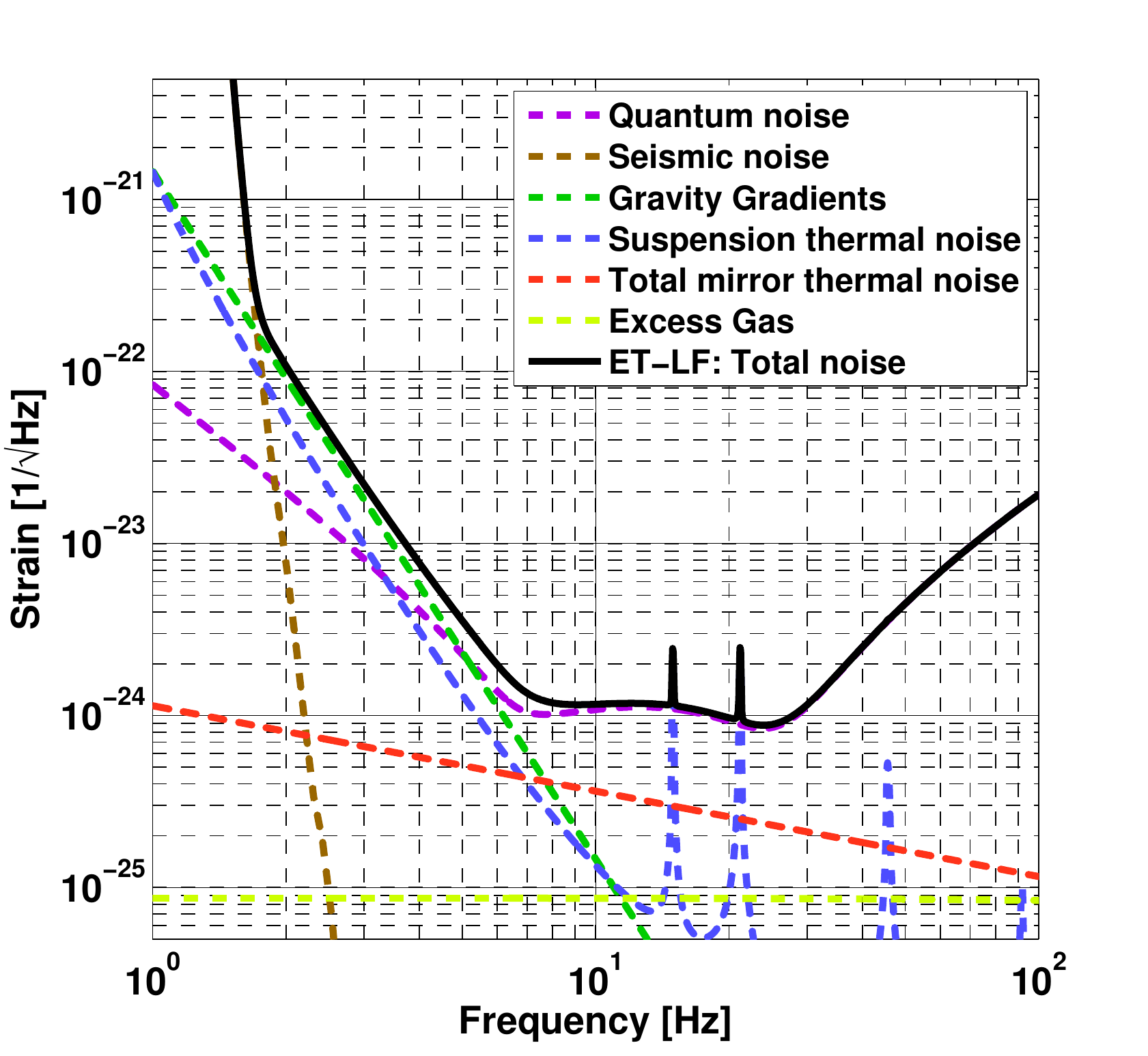}
\includegraphics[width=0.48\textwidth]{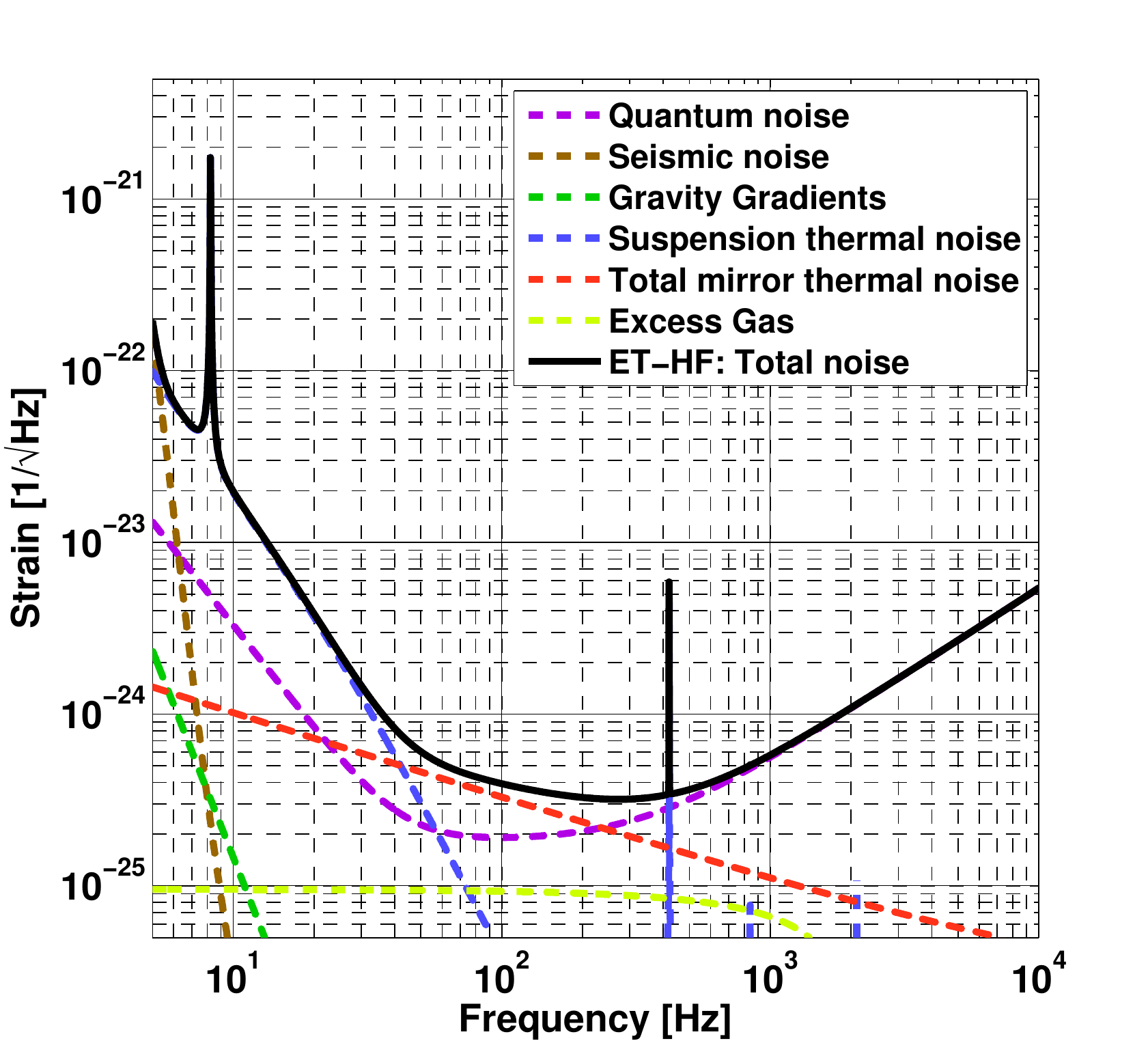}
\caption{Noise budgets for the ET-D low and high-frequency interferometers, using the parameters
given in Table \ref{tab:summary}. }\label{fig:ETD_budget}
\end{figure}

Using silicon test masses (62\,cm diameter, 30\,cm thickness) of 10\,K, a Young's modulus of silicon of 164\,GPa,
loss angles of $5 \times 10^{-5}$ and $2 \times 10^{-4}$ for the low and high-refraction 
coating materials\footnote{Unfortunately the available measurements indicate  higher
loss angles for the coating materials at cryogenic temperatures than at room
temperature \cite{Martin08}. However, since research on cryogenic coatings just started,
  we optimistically assumed   that by
the time construction of third-generation instruments starts, coatings will be 
available featuring the same loss angles as current coatings at room temperature
\cite{Harry06, harry07}}, respectively, and a laser  beam radius of 12\,cm, we get
a total mirror thermal noise contribution for the low-frequency 
detector as indicated by the orange dashed line in the right hand plot of Figure \ref{fig:SR_options}.
In this scenario the mirror thermal noise  would be  at least 
a factor of three below the quantum noise for all frequencies. Hence, we could in principle consider
reducing the beam size on the test masses, which could allow for the reduction of
the mirror size. Smaller test masses and smaller laser beams would be
advantageous for many aspects of the observatory design, such as for instance the
total mass of the cryogenic payload or the mode matching into the arm cavities. 
However, on the other hand reducing the mirror mass will  increase the radiation 
pressure noise contribution.

The right hand plot of Figure \ref{fig:SR_options} shows a trade-off analysis of
beam size and mirror mass. Starting from a mirror of 62\,cm diameter and 30\,cm
thickness which corresponds to a beam radius of 12\,cm, we reduce the mirror 
diameter to 52\,cm and 45\,cm, while keeping the aspect ratio of the mirror substrate 
constant. Using this assumption, already a small reduction in the beam size will
increase the radiation pressure noise dramatically and subsequently spoil the sensitivity in 
the sub-10\,Hz band. Therefore, we assume for the low frequency interferometer
of ET-D a reduced beam radius of 9\,cm, corresponding to an effective test mass
diameter of 45\,cm, but at the same time keep the overall test mass weight at
about 200\,kg. 
 
\begin{table}
\begin{center}
\begin{tabular}{l l l}
\hline 
\hline
Parameter & ET-D-HF   & ET-D-LF \\
\hline
Arm length & 10\,km & 10\,km \\
Input power (after IMC) & 500\,W & 3\,W \\
Arm power & 3\,MW & 18\,kW\\
Temperature & 290\,K &  10\,K  \\
Mirror material & Fused silica & Silicon \\ 
Mirror diameter / thickness & 62\,cm / 30\,cm & min 45\,cm/ TBD \\
Mirror masses & 200\,kg & 211\,kg \\
Laser wavelength & 1064\,nm & 1550\,nm \\
SR-phase & tuned (0.0) & detuned (0.6)\\
SR transmittance & 10\,\% & 20\,\% \\
Quantum noise suppression &  freq. dep. squeez.& freq. dep. squeez. \\
Filter cavities & $1 \times 10\,$km  & $2 \times 10\,$km\\
Squeezing level  & 10 dB (effective) & 10 dB (effective) \\
Beam shape &  LG$_{33}$& TEM$_{00}$\\
Beam radius & 7.25\,cm & 9\,cm \\
Scatter loss per surface & 37.5\,ppm & 37.5\,ppm \\
Partial pressurefor H$_2$O, H$_2$, N$_2$ & $10^{-8}$, $5\cdot10^{-8}$, $10^{-9}$  Pa &$10^{-8}$, $5\cdot10^{-8}$, $10^{-9}$  Pa  \\
Seismic isolation & SA, 8\,m tall & mod SA, 17\,m tall \\
Seismic (for $f>1$\,Hz) & $5\cdot 10^{-10}\,{\rm m}/f^2$ & $5\cdot 10^{-10}\,{\rm m}/f^2$  \\
Gravity gradient subtraction & none & none \\
\hline
\hline
\end{tabular}
\caption{Summary of the most important parameters of the ET-D high and low-frequency
interferometers as shown in Figure \ref{fig:ETD_budget}. SA = super attenuator,  freq. dep. squeez. = 
squeezing with frequency dependent angle.\label{tab:summary}}
\end{center}
\end{table}

\section{Overall sensitivity of the Einstein Telescope}\label{sec:ETD}

Table  \ref{tab:summary} shows the most important parameters of the ET-D
interferometers. The corresponding noise budgets for the high and low-frequency
interferometers are shown in figure \ref{fig:ETD_budget}. 
The sensitivity of the low-frequency detector is limited by seismic noise below
1.7\,Hz, while gravity gradient noise directly limits in the frequency band
between 1.7 and 6\,Hz. For all frequencies above 6\,Hz (apart from the violin
mode resonances) quantum noise is the limiting noise source.

 The crossover frequency of 
the sensitivities of the low and high-frequency interferometers is at about
35\,Hz. Above this frequency the high frequency interferometer is 
limited only by 2 noise sources: Mirror thermal noise limits the sensitivity between 
40 and 200\,Hz, while the high-frequency section is  limited by quantum
noise.

\begin{figure}[Htb]
\centering
\includegraphics[width=0.85\textwidth]{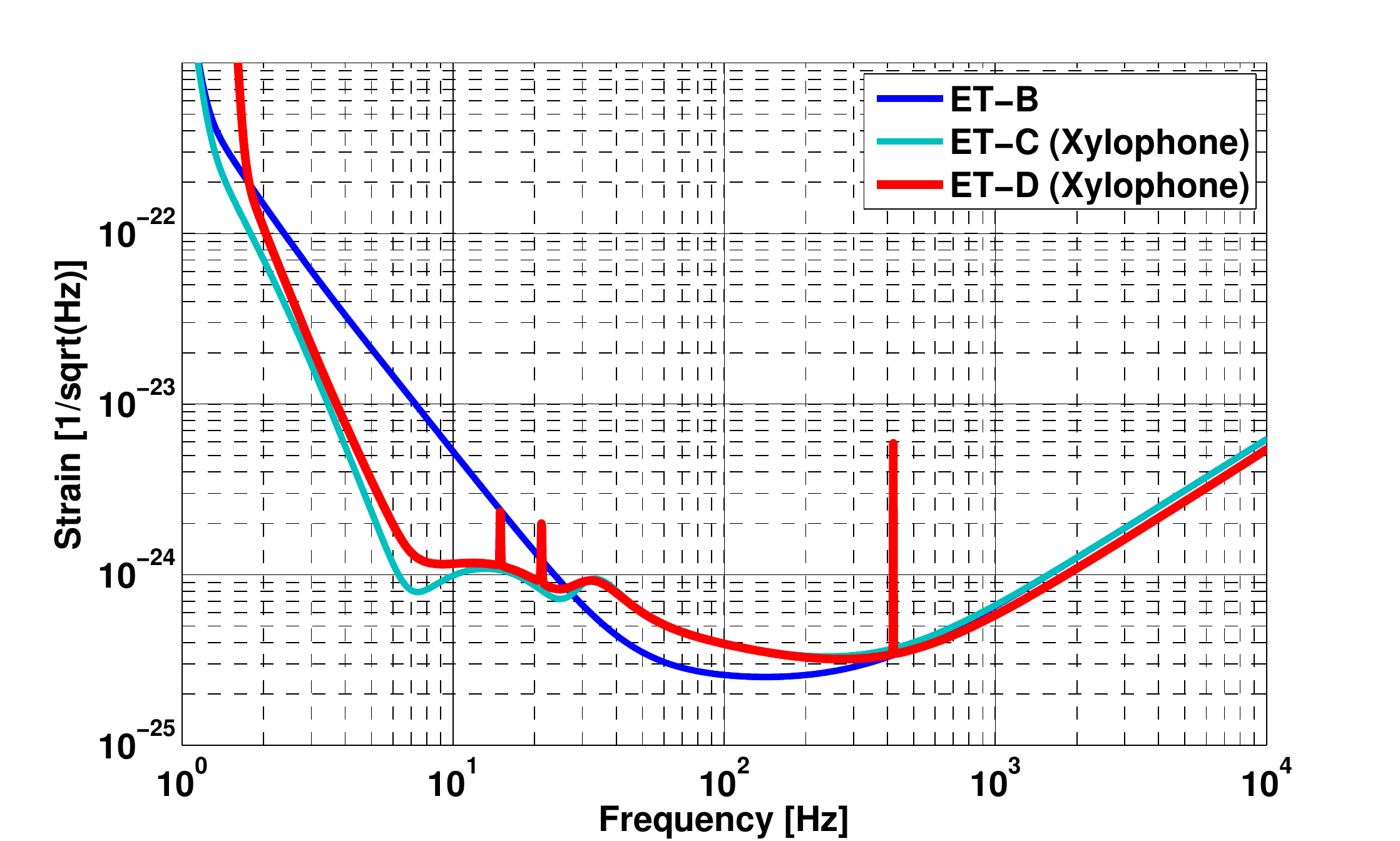}
\caption{Historical evolution of sensitivity models for the Einstein Telescope, starting from a single 
cryogenic broadband detector (ET-B) \cite{Hild08}, over the initial xylophone design (ET-C) \cite{Hild10} to the 
ET-D sensitivity described in this article.} \label{fig:all_sens}
\end{figure}

Figure \ref{fig:all_sens} shows the evolution of the sensitivity models for the
Einstein Telescope over the past years. The very first strawman design was 
based on a single cryogenic  interferometer covering the full frequency range of
interest (ET-B) \cite{Hild08}.  The introduction of the xylophone design
resulted in the ET-C sensitivity. In this article we significantly refined the
xylophone concept and obtained the ET-D sensitivity, which is slightly worse
than the ET-C sensitivity, but much more realistic. The loss of sensitivity
below 1.7\,Hz can be attributed to the application of a more realistic suspension
model yielding increased seismic noise. The slightly worse sensitivity in the
range between 2 and 8\,Hz is the result of the inclusion of suspension thermal
noise
for the cryogenic interferometer as well as the omission of any potential gravity gradient subtraction. 
The most significant difference between ET-C and ET-D shows up between 6 and 
10\,Hz and originates from including realistic squeezing losses experienced inside the filter cavities.

\section{Building a Full Third-Generation Observatory}\label{sec:triangle}

As discussed in the previous sections, one ET detector, covering the full
detection band will be made of two individual interferometers, one for 
low frequencies and one for high frequencies. However, the full ET observatory
will consist of 3 detectors arranged in triangular shape \cite{Freise09}, thus 
ultimately 6 interferometers will form the whole observatory. 

\begin{figure}[Htb]
\centering
\includegraphics[width=0.9\textwidth]{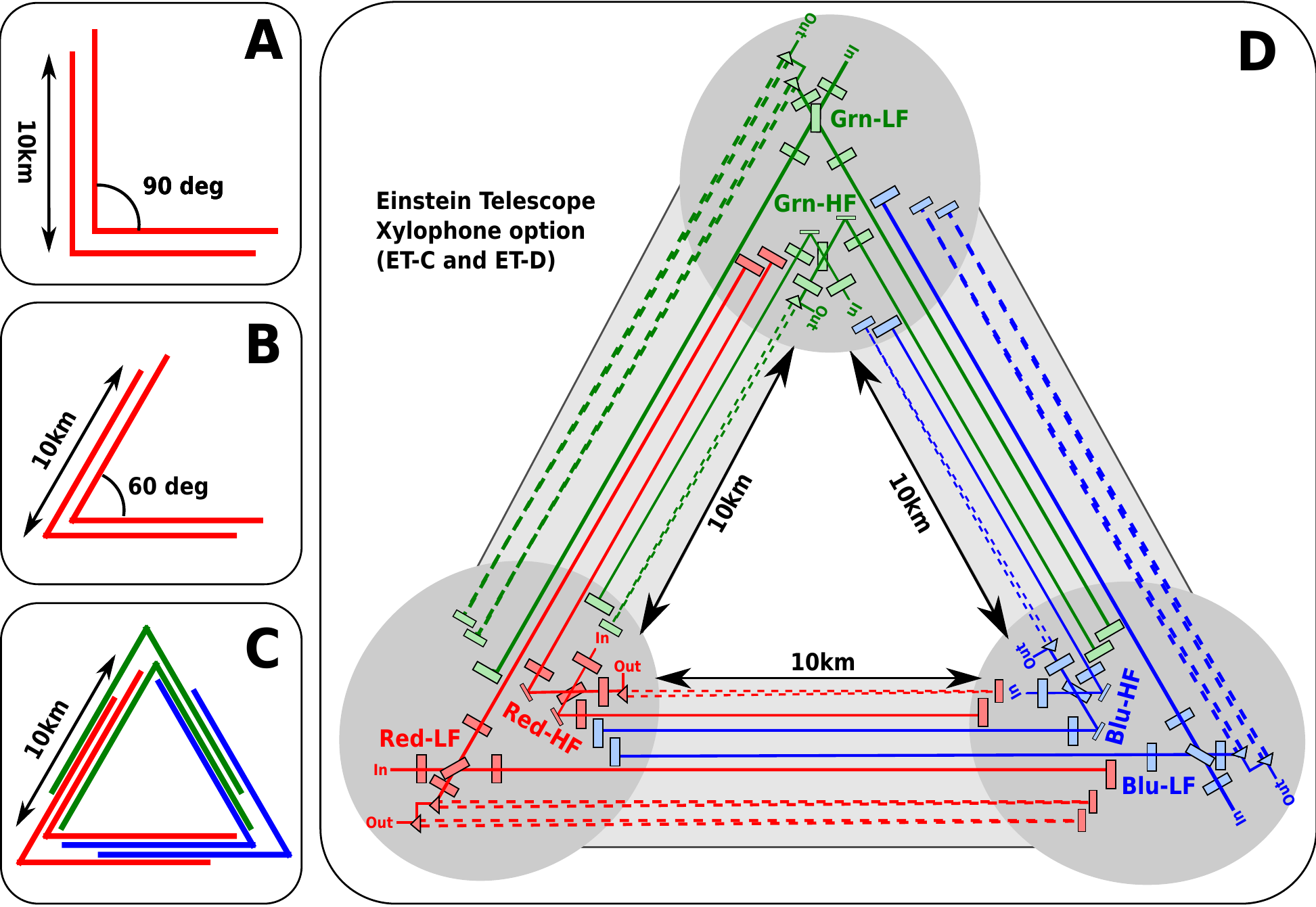}
\caption{Different interferometer configurations considered in this article. All sensitivities shown
in this paper refer to a pair of low and high-frequency interferometers forming a single detector of 
10\,km arm length and an opening angle of $90^\circ$. However, the full ET observatory 
will consist of 3 detectors with $60^\circ$ opening angle and arranged in the shape of
a triangle. Solid lines represent the main laser beams, while dashed lines indicate squeezed light beams.   } \label{fig:ET_triangle}
\end{figure}

It needs to be pointed out that the sensitivities described in this article
refer to a single pair of low and high frequency interferometers of 10\,km
arm length and an opening angle of $90^\circ$ (as shown in subplot A of figure
\ref{fig:ET_triangle}). 
The actual effective sensitivity of the full triangular ET observatory 
depends on the the orientation and polarisation of the source of interest. 
Let us assume a source directly positioned above the observatory emitting 
GW of plus polarisation. In case of  the configuration A in figure
\ref{fig:ET_triangle},
the sensitivity, $h(f)_{90}$, is then exactly represented by the ET-D trace in figure~\ref{fig:all_sens}. 

If we now decrease the opening angle of the two interferometers to $60^\circ$
(see configuration B in figure \ref{fig:ET_triangle}), 
the effective sensitivity for plus polarised GW is given by
\begin{equation}
h(f)_{60} = \frac{1}{\sin(60^\circ)} \times h(f)_{90} = 1.155 \times h(f)_{90},
\end{equation}
which is equivalent to shifting the ET-D curve in figure \ref{fig:all_sens} up by
about 15\,\%. 

Finally, if we consider the full triangle for a plus polarised source, we find
that the blue and red detectors in configuration C of figure \ref{fig:ET_triangle}
have the same sensitivity, while no signal shows up in the green detector. 
If we combine the red and blue detectors, the noise of the two needs to be added
in quadrature because it is uncorrelated, while the coherent signals need to be added up linearly. 
Thus the overall sensitivity of the full triangular observatory can be written as    
\begin{equation}
h(f)_{\Delta} =  \frac{1}{\sqrt{(\sin(60^\circ))^2+(\sin(60^\circ))^2}} \times h(f)_{90} =
0.816  \times h(f)_{90} ,
\end{equation}
which would be equivalent to shifting the ET-D curve in figure \ref{fig:all_sens} down by
about 18\,\%. 

Subplot D of figure \ref{fig:ET_triangle} shows a schematic drawing of the full
ET observatory configuration assumed for the ET-D sensitivity. Included are all
main mirrors of the 6 interferometers, as well as a total of 9 required
filter cavities for the frequency dependent squeezing. In total this sums up to  7 laser beams
per tunnel. 

\section{Summary and Future Plans}\label{sec:summary}

In this article we described a snapshot of the ongoing sensitivity studies for a
European third-generation GW observatory. The ET-D sensitivity represents a much
more realistic sensitivity compared to previous models, because we included new noise sources 
 as well as improved the accuracy of several already  previously included fundamental
noise sources. Key points of the new sensitivity model are the inclusion of
suspension thermal noise and a realistic seismic isolation system for the low
frequency interferometer and the proper accounting for squeezing losses inside the
filter cavities. Finally it needs to be pointed out that the current model does
not rely on any subtraction techniques for gravity gradient noise. 

In the future, we plan to further refine our sensitivity models by including noise
contributions from optical components outside the arm and filter cavities as well as by taking 
technical contributions such as laser frequency and laser amplitude noise into account.

\ack{The ET-D sensitivity data are available at {\tt http://www.et-gw.eu/etsensitivities}.
The authors are grateful for support from their hosting institutions and national funding 
agencies. 
This work has been performed with the support of the European Commission under the Framework
Programme 7 (FP7) `Capacities', project \emph{Einstein Telescope} (ET) design study 
(Grant Agreement 211743) (\tt http://www.et-gw.eu/)}

\end{document}